\documentclass{article}

\usepackage{amssymb}

\newtheorem{claim}{Claim}[section]
\newtheorem{theorem}[claim]{Theorem}
\newtheorem{proposition}[claim]{Proposition}
\newtheorem{lemma}[claim]{Lemma}
\newtheorem{remark}[claim]{Remark}

\newenvironment{proof}[1][Proof]{\textsl{#1:} }{\ \rule{0.4em}{0.7em}}

\begin{document}

%%%%%%%%%%%%%%%%%%%%%%%%%%%%%%%%%%%%%%%%%%%%%%%%%%%%%%%%%%%%%%
% title, author(s) and address(es) put here:                 %
%%%%%%%%%%%%%%%%%%%%%%%%%%%%%%%%%%%%%%%%%%%%%%%%%%%%%%%%%%%%%%

\title{Leaky quantum wire and dots: a resonance model}

\author{Pavel Exner and Sylwia Kondej}
\date{Department of Theoretical Physics, NPI, Academy of Sciences,
25068 \v{R}e\v{z}--Prague, Czechia; \\ [1mm] Institute of Physics,
University of Zielona G\'{o}ra, ul. Szafrana 4a, 65246 Zielona
G\'{o}ra, Poland}


\maketitle

\begin{abstract} \noindent We discuss a model of a leaky quantum wire and a
family of quantum dots described by Laplacian in
$L^2(\mathbb{R}^2)$ with an attractive singular perturbation
supported by a line and a finite number of points. The discrete
spectrum is shown to be nonempty, and furthermore, the resonance
problem can be explicitly solved in this setting; by
Birman-Schwinger method it is reformulated into a Friedrichs-type
model. \end{abstract}

%%%%%%%%%%%%%%%%%%%%%%%%%%%%%%%%%%%%%%%%%%%%%%%%%%%%%%%%%%%%%
% The main text of your paper                                         %
%%%%%%%%%%%%%%%%%%%%%%%%%%%%%%%%%%%%%%%%%%%%%%%%%%%%%%%%%%%%%

\section{Introduction}

In this talk we are going to discuss a simple model with the
Hamiltonian which is a generalized Schr\"odinger operator in
$L^2(\mathbb{R}^2)$. The interaction is supposed to be supported
by a line and a finite family of points, i.e. formally we have
 %----------------%
 \begin{equation} \label{forHam}
 -\Delta -\alpha \delta (x-\Sigma ) +\sum_{i=1}^n \tilde\beta_i
 \delta(x-y^{(i)})\,,
 \end{equation}
 %-----------
where $\alpha>0$, $\,\Sigma :=\{(x_1,0);\,x_1\in\mathbb{R}\}$, and
$y^{(i)}\in \mathbb{R}^2 \setminus \Sigma$; coupling constants of
the two-dimensional $\delta$ potentials will be specified below.
First one has to say a few words about a motivation of this
problem. Operators of the type (\ref{forHam}) or similar have been
studied recently as models of nanostructures which are ``leaky''
in the sense that they do not neglect quantum tunneling. While
various results about the discrete spectrum were derived
%\cite{e1,e2,ei,ek1,ek2,ek3,en1,en2,et,ey1,ey2,ey3,ey4}
\cite{e1}-\cite{ek3}, \cite{en1}-\cite{ey4}, much less is known
about scattering in this setting, in particular, about resonances.

The simple form of the interaction support, $\Sigma \cup\Pi$ with
$\Pi:= \{y^{(i)}\}$, will allow us to answer this question for the
operator (\ref{forHam}). We will achieve that by using the
generalized Birman-Schwinger method which makes it possible to
convert the original PDE problem into a simpler equation which in
the present situation is in part integral, in part algebraic. What
is important is that the method works not only for the discrete
spectrum but it can be used also to find singularities of the
analytically continued resolvent. The problem then boils down to a
finite rank perturbation of eigenvalues embedded in the continuous
spectrum, i.e. something which calls to mind the celebrated
Friedrichs model. To fit into the prescribed volume limit we
present here the main results with sketches of the proofs leaving
detailed arguments and extensions to a forthcoming paper
\cite{ek4}.

\section{The Hamiltonian and its resolvent}

A proper way to define (\ref{forHam}) as a self-adjoint operator
is through boundary conditions \cite{aghh}. Consider functions
$f\in W^{2,2}_{\mathrm{loc}}(\mathbb{R}^{2}\setminus (\Sigma \cup
\Pi)) \cap L^2(\mathbb{R}^2)$ which are continuous on $\Sigma $.
For small enough $r_i>0$ the restriction
$f\!\upharpoonright_{r_i}$ of $f$ to the circle
$\{x\in\mathbb{R}^2:\: |x-y^{(i)}|=r_i\}$ makes then sense. We say
that such an $f$ belongs to $D(\dot{H}_{\alpha, \beta})$ iff the
following limits
 %----------------%
 $$ \Xi_i(f):=-\lim_{r_i\to 0}\frac{1}{\ln r_i} f \!\upharpoonright
 _{r_i}\,,\quad \Omega_i(f):=\lim_{r_i\to 0}[f\!\upharpoonright
 _{r_i} +\Xi_i(f)\ln r_i] $$
 %----------------%
for $i=1,\dots,n$, and
 %----------------%
 $$ \Xi_{\Sigma}(f)(x_1):=\frac{\partial f}{\partial x_2}(x_1,0+) -
 \frac{\partial f}{\partial x_2}(x_1,0-)\,, \quad
 \Omega_{\Sigma}(f)(x_1):= f(x_1,0) $$
 %---------------%
are finite and satisfy the relations
 %----------------%
 \begin{equation}  \label{boucon}
 2\pi \beta_i \Xi_i(f)=\Omega  _i(f)\,, \quad \Xi
 _{\Sigma}(f)(x_1)=-\alpha \Omega  _{\Sigma }(f)(x_1)\,;
 \end{equation}
 %----------------%
we denote $\beta := (\beta_1,...,\beta_n)$. Then we define $\dot
H_{\alpha, \beta}:\:D(\dot H_{\alpha, \beta})\to
L^2(\mathbb{R}^2)$ acting as
 %----------------%
 $$ \dot H_{\alpha, \beta}f(x)= -\Delta f(x) \quad
 \mathrm{for}\quad x\in \mathbb{R}^{2} \setminus (\Sigma
\cup \Pi)\,, $$
 %----------------%
and $H_{\alpha, \beta}$ as its closure. Modifying the argument of
\cite{aghh} to the present situation one can check that
$H_{\alpha, \beta}$ is self-adjoint; an alternative way is to use
the method of \cite{p1}. We identify it with the formal operator
(\ref{forHam}). Notice that the $\beta_i$'s do not coincide with
the formal coupling constants in (\ref{forHam}), for instance,
absence of the point interaction at $y^{(i)}$ means
$\beta_i=\infty$.

The key element in spectral analysis of $H_{\alpha, \beta}$ is
finding an expression for its resolvent. Given $z\in
\mathbb{C}\setminus [0,\infty)$ we denote by $R(z):=(-\Delta
-z)^{-1}$ the free resolvent, which is an integral operator in
$L^{2}\equiv L^{2}(\mathbb{R}^{2})$ with the kernel
$G_{z}(x,x')=\frac{1}{2\pi}K_0 (\sqrt{-z}|x-x'|)$, where
$K_0(\cdot)$ is the Macdonald function and $z\mapsto \sqrt{z}$ has
a cut on the positive halfline. We also denote by $\mathbf{R}(z)$
the unitary operator defined as $R(z)$ but acting from $L^{2}$ to
$W^{2,2}\equiv W^{2,2}(\mathbb{R}^{2})$. To express the resolvent
of $H_{\alpha, \beta}$ we need two auxiliary Hilbert spaces,
$\mathcal{H}_0:= L^{2}(\mathbb{R})$ and $\mathcal{H}_1:=
\mathbb{C}^n$, and the corresponding trace maps $\tau
_0:W^{2,2}\to \mathcal{H}_0$ and $\tau _1 : W^{2,2}\to
\mathcal{H}_1$ which act as
 %---------------------%
 $$
 \tau _0  f:=f\!\upharpoonright_{\,\Sigma }\,, \quad \tau_1
 f:=f\!\upharpoonright_{\,\Pi}=(f\!\upharpoonright _{\,\{y^{(1)}\}},...,
 f\!\upharpoonright _{\,\{y^{(n)}\}})\,,
 $$
 %--------------------%
respectively; as before the used symbols means appropriate
restrictions. These maps in turn allow us to define canonical
embeddings of $\mathbf{R}(z)$ to $\mathcal{H}_i$ by
 %---------------------%
 \begin{equation} \label{embedd1}
 \mathbf{R}_{i,L}(z)=\tau _iR(z):\:L^{2}\to \mathcal{H}_i\,, \quad
 \mathbf{R}_{L,i}(z)=[\mathbf{R}_{i,L}(z)]^{\ast}:\:\mathcal{H}
 _i\to L^{2}
 \end{equation}
 %--------------%
and
 %--------------%
 \begin{equation} \label{embedd2}
 \mathbf{R}_{j,i}(z)=\tau _{j}\mathbf{R}_{L,i}(z):\mathcal{H}
 _i\to \mathcal{H}_{j}\,.
 \end{equation}
 %--------------%
We introduce the operator-valued matrix $ \Gamma(z)= [\Gamma
_{ij}(z)]:\: \mathcal{H}_0\oplus\mathcal{H}_1 \to
\mathcal{H}_0\oplus\mathcal{H}_1$ with the ``blocks''
$\Gamma_{ij}(z):\: \mathcal{H}_{j}\to \mathcal{H}_i$ given by
 %--------------%
 \begin{eqnarray*} \label{forg22}
 \Gamma _{ij}(z)g &=& -\mathbf{R}_{i,j}(z)g \qquad
 \mathrm{for}\quad i\neq j \quad \mathrm{and }\quad g\in
 \mathcal{H}_{j}\,, \\ \Gamma_{00}(z)f &=& \left[\alpha^{-1}
 -\mathbf{R}_{0,0}(z)\right] f \qquad \mathrm{if} \quad
 f\in \mathcal{H}_0\,, \\ \Gamma _{11}(z)\varphi &=&
 \left( s_{\beta }(z) \delta_{kl} - G_{z}(y^{(k)},y^{(l)}) (1\!-\!
 \delta_{kl}) \right) \varphi \qquad \mathrm{for} \quad \varphi \in
 \mathcal{H}_1\,,
 \end{eqnarray*}
 %--------------%
where $s_{\beta}(z)= \beta+s(z):=\beta +\frac{1}{2\pi}(\ln
\frac{\sqrt{z}} {2i}-\psi(1))$ and the operator in the last row is
written explicitly through the components of the corresponding
$n\times n$ matrix.

We will see that $\rho(H_{\alpha,\beta })$ coincides with the set
of $z$ for which $\Gamma(z)$ has a bounded inverse. The latter is
contained in $\mathbb{C}\setminus [-\frac{1}{4}\alpha^2, \infty)$,
hence we can define the ``reduced determinant''
 %--------------%
 $$
 D(z):=\Gamma_{11}(z)-\Gamma_{10}(z)
 \Gamma_{00}(z)^{-1}\Gamma _{01}(z)\::\: \mathcal{H}_1\to
 \mathcal{H}_1\,, $$
 %--------------%
by means of which the ``blocks'' of $[\Gamma(z)]^{-1}:\:
\mathcal{H}_0\oplus\mathcal{H}_1\to \mathcal{H}_0
\oplus\mathcal{H}_1$ express as
 %--------------%
 \begin{eqnarray*}
 \left[\Gamma(z)\right]_{11}^{-1} &=& D(z)^{-1}\,, \\
 \left[\Gamma(z)\right]_{00}^{-1} &=&
 \Gamma_{10}(z)^{-1} \Gamma_{11}(z)D(z)^{-1}
 \Gamma_{10}(z)\Gamma_{00}(z)^{-1}\,, \\
 \left[\Gamma(z)\right]_{01}^{-1} &=& -\Gamma_{00}(z)^{-1}
 \Gamma _{01}(z) D(z)^{-1}\,, \\
 \left[\Gamma(z)\right]_{10}^{-1} &=& -D(z)^{-1}
 \Gamma_{10}(z)\Gamma_{00}(z)^{-1}\,;
\end{eqnarray*}
 %--------------%
we use the natural notation which distinguishes them from the
inverses of $\Gamma_{ij}(z)$. Now we can state the sought
resolvent formula.

%---------------------------------------------------------------%
\begin{theorem} \label{resoth}
For $z\in \rho(H_{\alpha,\beta })$ with $\mathrm{Im\,}z>0$ the
resolvent of $H_{\alpha,\beta }$ is given by
 %------------%
 \begin{equation} \label{resol}
  R_{\alpha,\beta }(z)\equiv (H_{\alpha,\beta}-z)^{-1}
 =R(z) \,+\sum_{i,j=0}^{1} \mathbf{R}_{L,i}(z)[\Gamma(z)]_{ij}^{-1}
 \mathbf{R}_{j,L}(z)\,. \end{equation}
 %-------------%
\end{theorem}
%---------------------------------------------------------------%
\begin{proof}
For simplicity we will assume $n=1$ only, i.e. $\Pi=\{y\}\,$;
extension to the general case is easy. We have to check that $f\in
D(H_{\alpha,\beta})$ holds if and only if $f=\tilde
R_{\alpha,\beta}(z)g$ for some $g\in L^2$, where $\tilde
R_{\alpha,\beta}(z)$ denotes the operator at the right-hand side
of the last equation. Suppose that $f$ is of this form. It belongs
obviously to $W_{\mathrm{loc}}^{2,2}(\mathbb{R}^{2}\setminus
(\Sigma \cup \Pi)) \cap L^2$ because all its components belong to
this set. Combining the definitions of $\mathbf{R}_{i,j},\:
[\Gamma(z)]_{ij}^{-1}$, and functionals $\Xi(f)\equiv \Xi_1(f)$,
$\Omega_1(f)\equiv\Omega_1(f)$ introduced above with the
asymptotic behaviour of Macdonald function, $K_0(\sqrt{-z}\rho )=
-2\ln \rho -4\pi s(z)+\mathcal{O}(\rho)$ as $\rho\to 0$, we arrive
at
 %--------------%
 \begin{eqnarray*}
 2\pi \Xi (f) &=& \sum_{i=0}^1\, [\Gamma(z)]_{1i}^{-1}
 \mathbf{R}_{i,L}(z)g\,, \\
 \Omega (f) &=& \mathbf{R}_{1,L}(z)g -\sum_{i=0}^1\,
 \Gamma_{10}(z) [\Gamma(z)]_{0i}^{-1}\mathbf{R}_{i,L}g-
 s(z)\sum_{i=0}^1\, [\Gamma(z)]_{1i}^{-1}\mathbf{R}_{i,L}(z)g\,.
 \end{eqnarray*}
 %--------------%
Let us consider separately the components of $\Xi(f),\, \Omega
(f)$ coming from the behaviour of $g$ at the point $y$ and on
$\Sigma $, i.e. $\Xi^i(f):=\frac{1}{2\pi} [\Gamma(z)]_{1i}^{-1}
\mathbf{R}_{i,L}g$ and
 %--------------%
 \begin{eqnarray*}
 \Omega^0(f) &:=& \left[- \Gamma_{10}(z)[\Gamma(z)]_{00}^{-1}
 -s(z)[\Gamma(z)]_{10}^{-1}\right] \mathbf{R}_{0,L}g\,, \\
 \Omega^1(f) &:=& \left[ 1-\Gamma_{10}(z)[\Gamma(z)]_{01}^{-1}
 -s(z) [\Gamma(z)]_{11}^{-1}\right] \mathbf{R}_{1,L}g\,;
 \end{eqnarray*}
 %--------------%
using the properties of $[\Gamma_{ij}(z)]$ and its inverse it is
straightforward to check that $\Omega^i(f)=2\pi \beta \Xi^i(f)$
holds for $i=0,1$. Similar calculations yield the relation
$\Xi_{\Sigma }(f)= -\alpha \Omega_{\Sigma }(f)$ which means that
$f$ belongs to $D(H_{\alpha,\beta })$, and the converse statement,
namely that any function from $D(H_{\alpha,\beta })$ admits a
representation of the form $f=\tilde R_{\alpha,\beta}(z)g$. To
conclude the proof, observe that for such a function $f\in
D(H_{\alpha,\beta })$ which vanishes on $ \Sigma \cup \Pi$ we have
$(-\Delta-z)f=g$. Consequently, $\tilde R_{\alpha,\beta}(z)=
R_{\alpha,\beta}(z)$ is the resolvent of the Laplace operator in
$L^2(\mathbb{R}^2)$ with the boundary conditions (\ref{boucon}).
\end{proof}

In a similar way one can compare $R_{\alpha,\beta}(z)$ to the
resolvent $R_\alpha(z)$ of the operator $H_\alpha$ with the point
interactions absent using the operators $\mathbf{R}_{\alpha;
1,L}(z),\, \mathbf{R}_{\alpha; L,1}(z)$ mapping between $L^2$ and
$\mathcal{H}_1$, and $\mathbf{R}_{\alpha; 1,1}(z)\equiv \Gamma
_{\alpha; 11}(z)$ on $\mathcal{H}_1$ defined in analogy with
(\ref{embedd1}) and (\ref{embedd2}); the latter is
 %-------------%
 $$ \Gamma _{\alpha; 11}(z)\varphi =
 \left( s^{(\alpha)}_{\beta,k}(z) \delta_{kl} -
 G^{(\alpha)}_z(y^{(k)}, y^{(l)}) (1\!-\! \delta_{kl}) \right)
 \varphi \qquad \mathrm{for} \;\varphi \in
 \mathcal{H}_1\,, $$
 %-------------%
where $s^{(\alpha)}_{\beta,k}(z):= \beta - \lim_{\eta\to 0} \left(
G^{(\alpha)}_z(y^{(k)}, y^{(k)}\!+\eta) + \frac{1}{2\pi} \ln|\eta|
\right)$ and $G^{(\alpha)}_z$ is the integral kernel of the
operator $R_\alpha(z)$. Using the standard Krein-formula argument
mimicking \cite{aghh} we find that the two resolvents differ by
$\mathbf{R}_{\alpha;L,1}(z)[\Gamma_{\alpha;11}(z)]^{-1}
\mathbf{R}_{\alpha;1,L}(z)$. This can be simplified further: we
have $R_{\alpha}(z)=R(z)+R_{L,0}(z)\Gamma_{00}(z)^{-1} R_{0,L}(z)$
for $z\in \rho(H_{\alpha})=\mathbb{C}\setminus [-\frac{1}{4}
\alpha^2, \infty)$, and taking into account the asymptotic
behaviour of Macdonald function we get
 %------------%
 $$ s^{(\alpha)}_{\beta,k}(z) =s_{\beta}(z)
 -(\mathbf{R}_{1,0}(z)\Gamma_{00}(z)^{-1}
 \mathbf{R}_{0,1}(z))_{kk}\,.$$
 %------------%
Taken together, these considerations mean that
$\Gamma_{\alpha;1,1}(z) =D(z)$, or in other words
%---------------------------------------------------------------%
\begin{proposition} \label{resoprop}
For $z\in \rho(H_{\alpha,\beta })$ with $\mathrm{Im\,}z>0$ the
resolvent of $H_{\alpha,\beta }$ is given by
 %------------%
 $$ R_{\alpha,\beta }(z)
 =R_{\alpha}(z) + \mathbf{R}_{\alpha;L,1}(z)
 D(z)^{-1} \mathbf{R}_{\alpha;1,L}(z)\,. $$
 %-------------%
\end{proposition}
%---------------------------------------------------------------%

%%%%%%%%%%%%%%%%%%%%%%%%%%%%%%%%%%%%%%%%%%%%%%%%%%%%%%%%%%%%%

\section{Spectral properties}

Before addressing our main question about resonances in this
model, let us describe spectral properties of $H_{\alpha,\beta}$.
The spectrum of $H_{\alpha}$ is found easily by separation of
variables; using Proposition~\ref{resoprop} in combination with
Weyl's theorem and \cite[Thm.~XIII.19]{rs} we find that
 %------------%
 $$ \sigma_{\mathrm{ess}}(H_{\alpha,\beta})=
 \sigma_{\mathrm{ac}}(H_{\alpha,\beta})=
 [-\frac{1}{4}\alpha^2,\infty )\,. $$
 %-------------%
Less trivial is the discrete spectrum. An efficient way to
determine it is provided by the generalized Birman-Schwinger
principle, which in view of Theorem~\ref{resoth} reads
 %-------------%
 \begin{eqnarray}
 z\in\sigma_{\mathrm{disc}}(H_{\alpha,\beta})
 \,&\Leftrightarrow&\, 0\in\sigma_{\mathrm{disc}}(\Gamma(z))\,, \quad
 \dim\ker\Gamma(z) =\dim \ker (H_{\alpha,\beta}\!-\!z)\,, \label{gBSev} \\
 H_{\alpha,\beta}\phi_z =z\phi_z \,&\Leftrightarrow&\,
 \phi_z=\sum_{i=0}^{1}\mathbf{R}_{L,i}(z) \eta_{i,z} \quad
 \mathrm{for} \; z\in\sigma_{\mathrm{disc}}(H_{\alpha,\beta})\,,
 \label{gBSef}
 \end{eqnarray}
 %-------------%
where $(\eta_{0,z},\eta_{1,z})\in \ker\Gamma(z)$ -- cf.~\cite{p1}.
Moreover, it is clear from the explicit form of $[\Gamma(z)]^{-1}$
that $0\in\sigma_{\mathrm{disc}}(\Gamma (z)) \Leftrightarrow
0\in\sigma_{\mathrm{disc}} (D(z))$; this reduces the task to an
algebraic problem.

Consider again the case $n=1$ with the point interaction placed at
$(0,a)$ with $a>0$. In absence of the line, the operator
$H_{0,\beta}$ has a single eigenvalue $\epsilon_{\beta
}=-4\mathrm{e}^{2(-2\pi \beta +\psi (1))}$; we will show that
$\sigma_{\mathrm{disc}} (H_{\alpha,\beta})$ is nonempty for any
$\alpha>0$. More specifically, we claim that
%---------------------------------------------------------------%
\begin{theorem} \label{1disc}
For any $\alpha >0$ and $\beta\in\mathbb{R}$ the operator
$H_{\alpha,\beta}$ has one isolated eigenvalue $-\kappa_a^2$ with
the eigenvector given in terms of the Fourier transform
 %--------------%
 $$ \mathrm{const}\: \int_{\mathbb{R}^{2}}\left(
 \frac{\mathrm{e}^{-ip_{2}a}} {2\pi}+\frac{\alpha
 \mathrm{e}^{-(p_1^{2}+ \kappa_a^{2})^{1/2}a }}
 {(2(p_1^{2}+\kappa_a^{2})^{1/2}-\alpha )}\right)
 \frac{\mathrm{e}^{ipx}}{p^{2}+\kappa_a^{2}}\, \mathrm{d}p\,,  $$
 %--------------%
where $p=(p_1,p_{2})$. The function $a\mapsto -\kappa _{a}^{2}$ is
continuously increasing in $(0,\infty)$ and satisfies
$\lim_{a\to\infty}(-\kappa_a^2)= \min\left\{ \epsilon_{\beta},\,
 -\frac{1}{4} \alpha^2 \right\}$, while the opposite limit
 $-\kappa _0^2:= \lim_{a\to 0}(-\kappa_a^2)$ is finite.
\end{theorem}
%---------------------------------------------------------------%
\begin{proof} One has to find $z$ for which $\ker D(\cdot)$ is
nontrivial. We put $z=-\kappa ^2$ with $\kappa
>0$ and introduce $\breve{D}(\kappa):=D(-\kappa ^{2})$, and similarly
for other quantities. By a straightforward calculation we find
that $\breve{D}(\kappa)$ acts as a multiplication by
$\breve{\gamma}_a(\kappa):=\breve{s}_{\beta}(\kappa)-
\breve{\phi}_a(\kappa)$, where
 %--------------%
 \begin{equation} \label{phiaka}
 \breve{\phi}_a(\kappa ) =\frac{\alpha}{4\pi}
 \int_{\mathbb{R}}\frac{\mathrm{e}^{-2(p^2+\kappa^2)^{1/2}
 a}} {(2(p^2+\kappa^2)^{1/2}-\alpha)(p^2
 +\kappa^2)^{1/2}}\, \mathrm{d}p
 \end{equation}
 %---------------%
and $\breve{s}_{\beta}(\kappa )=\frac{1}{2\pi }\left[ \ln
\frac{\kappa}{2} -\psi(1)\right]$. It is straightforward to check
that $\kappa \to \breve{\gamma}_a(\kappa )$ is continuous,
strictly increasing, and tends to $\pm\infty$ as $\kappa \to
\infty$ and $\kappa\to \frac{1}{2}\alpha+$, respectively. Hence
the equation $\breve{\gamma}_a(\kappa)=0$ has a unique solution
$\kappa_a$ in $(\frac{1}{2}\alpha,\infty)$. Evaluating
$\breve{\mathbf{R}}_{L,1}(\kappa)$ we get the eigenfunction from
(\ref{gBSef}). Moreover, using (\ref{phiaka}) we find that for a
fixed $\kappa $ the function $a\mapsto \breve{\phi}_a(\kappa)$ is
decreasing; combining this with the fact that
$\breve{s}_\beta(\cdot)$ is increasing we conclude that
$a\mapsto\kappa_a$ is decreasing. Next we employ the relation
$\lim_{a\to\infty} \breve{\phi}_a(\kappa)=0$, which is easily seen
to be valid pointwise; in combination with $\breve{s}_\beta
(\sqrt{-\epsilon_{\beta}})=0$ it yields the sought limit for
$a\to\infty$. To finish the proof, recall that (\ref{phiaka}) is
bounded from above by $\breve{\phi}_0 (\kappa)$ and the equation
$\breve{s}_\beta(\kappa)- \breve{\phi}_0(\kappa)=0$ has a unique
finite solution $\kappa_0$.
\end{proof}

If $n>1$ the structure of the spectrum becomes more complicated.
For instance, it is clear that $H_{\alpha,\beta }$ can have
embedded eigenvalues provided the sets $\Pi$ and $\beta$ have a
mirror symmetry w.r.t. $\Sigma$ and $\sigma_{\mathrm{disc}}
(H_{0,\beta}) \cap \left(-\frac{1}{4}\alpha^2,0 \right) \ne
\emptyset$. In this short paper we restrict ourselves to quoting
the following general result, referring to \cite{ek4} for proof
and more details.
%------------------------------------%
\begin{theorem} %\label{}
For any $\alpha>0$ and $\beta=(\beta_1,\dots,\beta_n)\subset
\mathbb{R}^n$ the operator $H_{\alpha,\beta}$ has $N$ isolated
eigenvalues, where $1\le N\le n$. In particular, if all the point
interactions are strong enough, i.e. the numbers $-\beta_i$ are
sufficiently large, we have $N=n$.
\end{theorem}

%%%%%%%%%%%%%%%%%%%%%%%%%%%%%%%%%%%%%%%%%%%%%%%%%%%%%%%%%%%%%

\section{Resonances}

\subsection{Poles of the continued resolvent}

For simplicity we consider again a single point interaction placed
at $y=(0,a)$ with $a>0$. In addition we have to assume that if the
tunneling between $y$ and the line is neglected, the point
interaction eigenvalue is embedded into the continuous spectrum of
$H_\alpha$, in other words, that $\epsilon_\beta>
-\frac{1}{4}\alpha^2$. As usual analyzing resonances means to
investigate singularities in the analytical continuation of
$R(\cdot)$ from the ``physical sheet'' across the cut
$[-\frac{1}{4}\alpha^2, \infty)$. Our main insight is that the
constituents of the operator at the right-hand side of
(\ref{resol}) can be separately continued analytically.
Consequently, one can extend the Birman-Schwinger principle to the
complex region and to look for zeros in the analytic continuation
of $D(\cdot)$. A direct calculation shows that $D(z)$ acts for
$z\in\mathbb{C} \setminus [-\frac{1}{4}\alpha^2, \infty)$ as a
multiplication by
 %--------------%
 \begin{equation} \label{resdet}
 d_a(z):= s_{\beta}(z)- \phi_a(z) = s_{\beta}(z)
 -\int_0^{\infty }\frac{\mu(z,t)}{t-z- \frac{1}{4}\alpha^2}\,
 \mathrm{d}t\,, \end{equation}
 %--------------%
where
 %--------------%
 $$ \mu(z,t):= \frac{i\alpha}{16\pi}\, \frac{(\alpha
 -2i(z\!-\!t)^{1/2})\, \mathrm{e}^{2ia(z-t)^{1/2}}}
 {t^{1/2}(z\!-\!t)^{1/2}}\,.$$
 %--------------%
We shall construct the continuation of $d_a$ to a region
$\Omega_-$ of the other sheet which has the interval
$(-\frac{1}{4}\alpha^2,0)$ as a part of its boundary at the real
axis. To this aim we need more notation. Put $\mu^{0}(\lambda,t):=
\lim_{\varepsilon\to 0} \mu(\lambda\!+\!i\varepsilon,t)$ and for
$\lambda \in (-\frac{1}{4}\alpha ^2,0)$ introduce the symbol
 %--------------%
 $$ I(\lambda):=\mathcal{P}\int_0^{\infty}
 \frac{\mu^{0}(\lambda,t)}{t-\lambda
 -\frac{1}{4}\alpha^2}\, \mathrm{d}t$$
 %--------------%
with the integral understood as its corresponding principal value.
Finally, we denote
 %--------------%
 $$ g_{\alpha ,a}(z):= \frac{i\alpha}{4}\,
 \frac{\mathrm{e}^{-\alpha a}}
 {(z +\frac{1}{4}\alpha^2)^{1/2}} \quad \mathrm{for}
 \; z\in\Omega_- \cup (-\frac{1}{4}\alpha ^2,0)\,. $$
 %--------------%
 %--------------------------------------%
\begin{lemma} \label{anacon}
The function $z\mapsto \phi _{a}(z)$ defined in (\ref{resdet}) can
be continued analytically across $(-\frac{1}{4}\alpha^2,0)$ to a
region $\Omega_-$ of the second sheet as follows,
 %--------------%
 \begin{eqnarray*}
 \phi_a^{0}(\lambda) =I(\lambda) +g_{\alpha,a}(\lambda )
 \quad &\mathrm{for}\quad& \lambda \in
 (-\frac{1}{4}\alpha^2,0)\,, \\
 \phi_a^{-}(z)= -\int_0^{\infty } \frac{\mu(z,t)}
 {t-z-\frac{1}{4}\alpha^2}\, \mathrm{d}t -2g_{\alpha,a}(z)
 \quad &\mathrm{for}\quad& z\in\Omega_-,\: \mathrm{Im\,}z<0\,.
 \end{eqnarray*}
 %--------------%
 \end{lemma}
%--------------------------------------%
\begin{proof} By a direct if tedious computation -- cf.~\cite{ek4}
-- one can verify the relations
 %--------------%
 $$
 \lim_{\varepsilon \to 0^{+}}\phi_a^{\pm}(\lambda \pm
 i\varepsilon ) =\phi_a^{0}(\lambda)\,,\qquad -\frac{1}{4}\alpha^2
 < \lambda <0\,, $$
 %--------------%
where $\phi ^{+}_{a}\equiv \phi _{a}$; so the claim of the lemma
follows from the edge-of-the-wedge theorem.
\end{proof}

Notice that apart of fixing a part of its boundary, we have
imposed no restrictions on the shape of $\Omega_-$. The lemma
allows us in turn to construct the analytic continuation of
$d_a(\cdot)$ across the same segment of the real axis. It is given
by the function $\eta_a: M\mapsto\mathbb{C}$, where
$M=\{z:\mathrm{Im\,}z>0\}\cup (-\frac{1}{4}\alpha^2,0)
\cup\Omega_-$ acting as
 %--------------%
 $$ \eta_a(z)=s_\beta(z)-\phi_a^{l(z)}(z)\,,$$
 %--------------%
where $l(z)=\pm$ if $\pm\mathrm{Im\,} z>0$ and $l(z)=0$ if
$z\in(-\frac{1}{4}\alpha^2,0)$, respectively. The problem at hand
is now to show that $\eta_a(\cdot)$ has a second-sheet zero, i.e.
for some $z\in \Omega _{-}$. To proceed further it is convenient
to put $\varsigma_{\beta} :=\sqrt{-\epsilon_{\beta}}$, and since
we are interested here primarily in large distances $a$, to make
the following reparametrization,
 %--------------%
 $$ b:=\mathrm{e}^{- a \varsigma _{\beta }}\quad \mathrm{and}\quad
 \tilde\eta(b,z):=\eta_a(z):\:[0,\infty) \times M\mapsto
 \mathbb{C}\,;
 $$
 %--------------%
we look then for zeros of the function $\tilde\eta$ for small
values of $b$. With this notation we have
 %--------------%
\begin{equation}\label{reparb}
\mu^{0}(\lambda,t) =\frac{\alpha }{16\pi }\frac{(\alpha
+2(t-\lambda )^{1/2})\, b^{2(t-\lambda )^{1/2}/\varsigma_{\beta}
}}{t^{1/2}(t-\lambda )^{1/2}}\,, \quad g_{\alpha ,a(b)}(\lambda)=
\frac{i\alpha}{4}\, \frac{b^{\alpha /\varsigma _{\beta }}}
 {(\lambda +\frac{1}{4}\alpha^2)^{1/2}}\,,
\end{equation}
 %--------------%
for $ \lambda \in (-\frac{1}{4}\alpha ^2,0)$, and similarly for
the other constituents of $\tilde\eta$. This yields our main
result.
%--------------------------------------%
\begin{theorem} \label{resonth}
Assume $\epsilon_{\beta }> -\frac{1}{4}\alpha^2$. For any $b$
small enough the function $\tilde\eta(\cdot,\cdot)$ has a zero at
a point $z(b)\in \Omega_-$ with the real and imaginary part,
$z(b)=\mu(b)+ i\nu(b),\; \nu(b)<0,$ which in the limit $b\to 0$,
i.e. $a\to\infty$, behave in the following way,
 %---------------%
 \begin{equation}\label{assyRI}
 \mu(b)=\epsilon_{\beta}+\mathcal{O}(b)\,, \quad \nu(b)
 =\mathcal{O}(b)\,.
 \end{equation}
 %---------------%
 \end{theorem}
 %---------------%
\begin{proof} By assumption we have $\varsigma _{\beta }\in
(0,{1\over 2}\alpha)$. Using formulae (\ref{reparb}) together with
the similar expressions of $\mu (z,t)$ and $g_{\alpha ,a}(z)$ in
terms of $b$ one can check that for a fixed $b\in [0,\infty )$ the
function $\tilde\eta (b,\cdot )$ is analytic in $M$ while with
respect to both variables $\tilde\eta $ is just of the $C^1$ class
in a neighbourhood of the point $(0,\epsilon_{\beta })$. Moreover,
it is easy to see that $\tilde\eta(0,\epsilon_{\beta})=0$ and
$\partial_z\tilde\eta(0,\epsilon_{\beta}) \neq 0$. Thus by the
implicit function theorem there exists a neighbourhood $U_{0}$ of
zero and a unique function $z(b):\:U_{0}\mapsto \mathbb{C}$ such
that $\tilde\eta(b,z(b))=0$ holds for all $b\in U_{0}$. Since
$H_{\alpha,\beta}$ is self-adjoint, $\nu(b)$ cannot be positive,
while $z(b)\in (-{1\over 4}\alpha ^2,0)$ for $b\neq 0$ can be
excluded by inspecting the explicit form of $\tilde\eta $.
Finally, by smoothness properties of $\tilde\eta$ both the real
and imaginary part of $z(b)$ are of the $C^1$ class which yields
the behaviour (\ref{assyRI}).
\end{proof}

\begin{remark} {\rm Since $\Omega_-$ can be arbitrarily extended to
the lower complex halfplane and all the quantities involved depend
analytically on $a$, it is natural to ask what happens with the
pole for other values of $a$. Using Lemma~\ref{anacon} one can
check that in the limit $\mathrm{Im\,}z\to -\infty$ we have
$|\phi^{-}_{a}(z)|\to 0$ uniformly in $a$ and $|s_{\beta}(z)|\to
\infty$. Thus the imaginary part of the solution $z(a)$ to
$s_{\beta }(z)-\phi^{-}_{a}(z)=0$ is bounded as a function of $a$,
and in particular, the resonance pole survives as $a\to 0$. On the
other hand, this argument says nothing about the residue.}
\end{remark}

\subsection{Scattering}

Let us consider now the same problem from the viewpoint of
scattering in the system $(H_{\alpha,\beta},H_\alpha)$. In view of
Proposition~\ref{resoprop} and Birman-Kuroda theorem the wave
operators exist and are complete; our aim is to find the on-shell
S-matrix in the interval $(-\frac{1}{4}\alpha^2,0)$, i.e. the
corresponding transmission and reflection amplitudes. Using the
notation introduced above and Proposition~\ref{resoprop} we can
write the resolvent for $\mathrm{Im\,}z>0$ as
 %--------------%
 $$ R_{\alpha ,\beta }(z)=R_{\alpha }(z) +\eta_a(z)^{-1}(\cdot,
 v_{z})v_{z}, $$
 %--------------%
where $v_{z}:=R_{\alpha;L,1}(z)$. We apply this operator to
$\omega_{\lambda +i\varepsilon}(x):= \mathrm{e}^{i(\lambda
+i\varepsilon +\alpha^2/4)^{1/2}x_1}\, \mathrm{e}^{-\alpha
|x_{2}|/2}$ and take the limit $\varepsilon\to 0+$ in the sense of
distributions; then a straightforward if tedious calculation shows
that $H_{\alpha,\beta}$ has a generalized eigenfunction which for
large $|x_1|$ behaves as
 %--------------%
 $$ \psi_\lambda(x)\approx \mathrm{e}^{i(\lambda+\alpha ^2/4)^{1/2}x_{1}}
 \, \mathrm{e}^{-\alpha|x_{2}|/2} +\frac{i}{4}\, \alpha
 \eta_a(\lambda)^{-1}\, \frac{\mathrm{e}^{-\alpha a}}{(\lambda
 +\frac{1}{4}\alpha^2)^{1/2}}\: \mathrm{e}^{i(\lambda
 +\alpha^2/4)^{1/2}|x_{1}| }e^{-\alpha|x_{2}|/2} $$
 %--------------%
for each $\lambda\in(-\frac{1}{4}\alpha^2,0)$. This yields the
sought quantities.
%---------------------------------------------------------------%
\begin{proposition} \label{scatres}
The reflection and transmission amplitudes are given by
 %------------%
 $$ \mathcal{R}(\lambda)= \mathcal{T}(\lambda)-1= \frac{i}{4}\,
 \alpha \eta_a(\lambda)^{-1}\, \frac{\mathrm{e}^{-\alpha a}}
 {(\lambda +\frac{1}{4}\alpha^2)^{1/2}}\,; $$
 %-------------%
they have the same pole in the analytical continuation to
$\Omega_-$ as the continued resolvent.
\end{proposition}
%---------------------------------------------------------------%

\subsection{Resonances induced by broken symmetry}

If $n\ge 2$ the resonance structure may become more complicated. A
new feature is the occurrence of resonances coming from a
violation of mirror symmetry. We will illustrate it on the
simplest example of a pair of point interactions placed at
$x_1=(0,a)$ and $x_2=(0,-a)$ with $a>0$ and coupling $\beta_b:=
(\beta,\beta+b)$, where $b$ is the symmetry-breaking parameter. We
choose $\alpha,\,a,\,\beta$ in such a way that the Hamiltonian
$H_{0,\beta_0}$ with two identical point interactions spaced by
$2a$ has two eigenvalues, the larger of which -- called
$\epsilon_2$ -- exceeds $-\frac{1}{4}\alpha^2$. As we have pointed
out, $H_{\alpha,\beta_0}$ has then in view of antisymmetry the
same eigenvalue $\epsilon_2$ embedded in the negative part of its
continuous spectrum.

Modifying the argument which led us to Theorem~\ref{resoth} we
have now to continue analytically the $2\times 2$ matrix $D(z)$
and find zeros of its determinant. This yields the equation
 %------------%
 \begin{equation} \label{violsym}
 s_{\beta}(z)(s_{\beta}(z)+b) -K_{0}(2a\sqrt{-z})^2 -
 (2s_{\beta}(z)+b)\phi^{l(z)}_a(z)- 2K_{0}(2a\sqrt{-z})
 \phi^{l(z)}_a(z) = 0\,, \end{equation}
 %-------------%
where $\phi ^{l(z)}_{a}(\cdot)$ is defined in Lemma~\ref{anacon}
and the left-hand side can be understood as a function
$\hat\eta(b,z):\,\mathbb{R} \setminus \{0\} \times M \to
\mathbb{C}$. We denote also $\kappa_2=\sqrt{-\epsilon_2}$,
$\tilde{g}(\lambda):= -ig_{\alpha,a}(\lambda)$ and put
$\breve{s}'_{\beta }(\cdot )$, $K_{0}'(\cdot )$ for corresponding
derivatives; then we have the following result.
%-----------------------------------------------------------%
\begin{theorem}
Suppose that $\epsilon_2\in(-\frac{1}{4}\alpha^2,0)$, then for all
nonzero $b$ small enough the equation (\ref{violsym}) has a
solution $z_2(b)\in \Omega_-$ with the real and imaginary part,
$z_2(b)=\mu_2(b)+ i\nu_2(b)$, which are real-analytic functions
with the following expansions,
 %---------------%
 \begin{eqnarray*}
 \mu_2(b) &=& \epsilon_2+ \frac{\kappa_2} {\breve{s}'_{\beta}
 (\kappa_2)+2aK_0'(2a\kappa_2)}\:b+ \mathcal{O}(b^2) \,, \\
 \nu_2(b) &=& -\frac{\kappa_2\tilde{g}(\epsilon_2)}
 {2(\breve{s}'_{\beta}(\kappa_2)+2aK_0'(2a\kappa_2))
 |\breve{s}_{\beta}(\kappa_2)-\phi_a^0(\epsilon_2)|^2}\:
 b^2 +\mathcal{O}(b^3)\,.
 \end{eqnarray*}
 %---------------%
\end{theorem}
%-----------------------------------------------------------%
\begin{proof} As in Theorem~\ref{resonth} we
rely on the implicit function theorem, but $\tilde\eta$ is now
jointly analytic, so is $z_2$. Since $\breve{s}'_{\beta}
(\kappa_2)+2aK_0'(2a\kappa_2)>0$ the leading term of $\nu_2(b)$ is
negative.
\end{proof}

%%%%%%%%%%%%%%%%%%%%%%%%%%%%%%%%%%%%%%%%%%%%%%%%%%%%%%%%%%%%%
% Doing Acknowledgement                                               %
%%%%%%%%%%%%%%%%%%%%%%%%%%%%%%%%%%%%%%%%%%%%%%%%%%%%%%%%%%%%%

\section*{Acknowledgments}

S.K. is grateful for the hospitality in Nuclear Physics Institute,
AS CR, where a part of this work was done. The research has been
partially supported by the GAAS Grant A1048101.

%%%%%%%%%%%%%%%%%%%%%%%%%%%%%%%%%%%%%%%%%%%%%%%%%%%%%%%%%%%%%
% Doing references:                                         %
%%%%%%%%%%%%%%%%%%%%%%%%%%%%%%%%%%%%%%%%%%%%%%%%%%%%%%%%%%%%%


\begin{thebibliography}{0}

%%%%%%%%%%%%%%%%%%%%%%%%%%%%%%%%%%%%%%%%%%%%%%%%%%%%%%%%%%%%%
%                                                           %
% Command to used is:-                                      %
%                                                           %
%  \bibitem{REFERENCE_LABEL} AUTHORS NAMES,                 %
%  {\it JOURNAL'S NAMES}{\bf VOLUME NUMBER}, PAGE (YEAR).   %
%                                                           %
%  See example below.                                         %
%                                                           %
%%%%%%%%%%%%%%%%%%%%%%%%%%%%%%%%%%%%%%%%%%%%%%%%%%%%%%%%%%%%%

\bibitem{aghh} S.~Albeverio, F.~Gesztesy, R.~H\o egh-Krohn, H.~Holden,
{\it Solvable Models in Quantum Mechanics}, Springer, Heidelberg
1988.

\bibitem{e1} P. Exner, {\em Lett. Math. Phys.} {\bf 57}, 87 (2001).

\bibitem{e2} P. Exner, in {\em Proceedings of the NSF Summer Research
Conference (Mt. Holyoke 2002)}; AMS ``Contemporary Mathematics"
Series, 2003.

\bibitem{ei} P. Exner and T. Ichinose, {\it J. Phys. A: Math. Gen.}
{\bf 34}, 1439 (2001).

\bibitem{ek1} P. Exner, S.~Kondej, {\em Ann. H.~Poincar\'{e}}
{\bf 3}, 967 (2002).

\bibitem{ek2} P. Exner, S.~Kondej, {\it J. Phys. A: Math. Gen.}
{\bf 36}, 443 (2003).

\bibitem{ek3} P. Exner, S.~Kondej, \texttt{math-ph/0303033}

\bibitem{ek4} P. Exner, S.~Kondej: Leaky quantum graphs: a solvable
resonance model, {\em in preparation}

\bibitem{en1} P. Exner, K.~N\v{e}mcov\'{a}, {\it J. Phys. A: Math. Gen.}
{\bf 34}, 7783 (2001).

\bibitem{en2} P. Exner, K.~N\v{e}mcov\'{a}, \texttt{math-ph/0306033}

\bibitem{et} P. Exner, M.~Tater, \texttt{math-ph/0303006}

\bibitem{ey1} P. Exner, K.~Yoshitomi, {\it J. Geom. Phys.}
{\bf 41}, 344 (2002).

\bibitem{ey2} P. Exner, K.~Yoshitomi, {\it Ann. H.~Poincar\'e}
{\bf 2}, 1139 (2001).

\bibitem{ey3} P. Exner, K.~Yoshitomi, {\it J. Phys. A: Math. Gen.}
{\bf 35}, 3479 (2002).

\bibitem{ey4} P. Exner, K.~Yoshitomi, {\em Lett. Math Phys.} (2003),
to appear; \texttt{math-ph/0303072}.

\bibitem{p1} A.~Posilicano, {\em J. Funct. Anal. } {\bf 183}, 109
(2001), and {\em Ann. Scuola Norm. Sup. Pisa}, to appear

\bibitem{rs} M.~Reed, B.~Simon,
{\it Methods of Modern Mathematical Physics IV}, Academic Press,
N.Y. 1978.


\end{thebibliography}
\end{document}